\documentclass[prx,reprint,english,superscriptaddress,floatfix]{revtex4-2}
\usepackage{graphicx}
\usepackage{bm}
\usepackage{amsmath}
\usepackage{amsfonts}
\usepackage{amsbsy}
\usepackage{amssymb}
\usepackage[mathscr]{eucal}
\usepackage{color}
\definecolor{trgcolor}{rgb}{0.80,0,0}

\usepackage{mathtools}
\begin{document}

\author{Todd R. Gingrich}
\email{todd.gingrich@northwestern.edu}
\affiliation{Department of Chemistry, Northwestern University, 2145 Sheridan Road, Evanston, Illinois 60208, USA}
\affiliation{Department of Engineering Sciences and Applied Mathematics, Northwestern University, 2145 Sheridan Road, Evanston, Illinois 60208, USA}
\thanks{All authors contributed equally to this work and are listed alphabetically.}

\author{Oleg A. Igoshin}
\email{igoshin@rice.edu}
\affiliation
{Center for Theoretical Biological Physics, Rice University, Houston TX 77005}
\affiliation
{Department of Chemistry, Rice University, Houston, TX 77005}
\affiliation
{Department of Bioengineering, Rice University, Houston, TX 77005}
\affiliation
{Department of Biosciences, Rice University, Houston, TX 77005}
\thanks{All authors contributed equally to this work and are listed alphabetically.}
\author{Anatoly B. Kolomeisky}
\email{tolya@rice.edu}
\affiliation
{Center for Theoretical Biological Physics, Rice University, Houston TX 77005}
\affiliation
{Department of Chemistry, Rice University, Houston, TX 77005}
\affiliation
{Department of Chemical and Biomolecular Engineering, Rice University, Houston, TX 77005}
\affiliation
{Department of Physics and Astronomy, Rice University, Houston, TX 77005}
\thanks{All authors contributed equally to this work and are listed alphabetically.}
\author{Dmitrii E. Makarov}
\email{makarov@cm.utexas.edu}
\affiliation{Department of Chemistry, University of Texas at Austin, Austin, TX, 78712 }
\affiliation{Oden Institute for Computational Engineering and Sciences, University of Texas at Austin, Austin, TX, 78712 }
\thanks{All authors contributed equally to this work and are listed alphabetically.}

\begin{abstract}
    Single-molecule or single-particle tracking measurements inherently yield noisy microscopic trajectories, often significantly constrained by the diffraction limit and by the finite rate at which photons are emitted and counted. Here we study systematically the resulting effects of finite spatial and temporal resolution on one's ability to discern and quantify the arrow of time in microscopic trajectories. Given an experimental time series $Y(t)$ degraded by noise, we consider the problem of estimating the entropy production associated with the corresponding microscopic variable $X(t)$ using two strategies. The first attempts to infer the statistical properties of $X(t)$ from those of $Y(t)$ before estimating the entropy production. The second uses the experimental observable as a proxy for the true microscopic observable, with the entropy production estimator applied directly to $Y(t)$. We prove that both strategies result in lower bounds on the true entropy production. Importantly, noise-degraded observables $Y(t)$ undergo non-Markovian dynamics even when $X(t)$ are Markovian, and non-Markovian entropy production estimators are advantageous. We further note nontrivial interplay between spatial and temporal resolution: in the presence of detection noise, improving the temporal resolution alone may lead to poorer rather than better entropy production estimates.
\end{abstract}
\title{Thermodynamic inference from noisy single-molecule time series}

\maketitle

\section{Introduction}

Differentiating between equilibrium and nonequilibrium active matter is often surprisingly difficult at a microscopic scale, where the directionality of the observed processes is masked by thermal motion, and where only partial microscopic information is available. During the past decade, many methods have been proposed that allow thermodynamic inference from partial observations or observations of coarse observables.\cite{peliti2021stochastic,SeifertBook,roldan2012entropy,hartich2021violation,Snippets,van2022thermodynamic,Polettini,li2019quantifying,Horowitz2019NP,Bisker2017,Gingrich2017JPAMT,PhysRevLett.116.120601,Harunari_2024} Yet most of these methods assume (1) infinite time resolution and (2) that observations yield ``true'', noiseless observables. These assumptions are almost never realized in experiments.\cite{challengesJPCL} For example, in single-molecule experiments, the temporal resolution is limited by the speed of the camera or by the rate at which individual photons arrive, and the spatial resolution is limited by diffraction. Only a handful of studies have explicitly considered the effects of experimental resolution on thermodynamic inference.\cite{Fritz2025,Bauer_2025,Song2024FRET,Yu_2024}

Here, we consider the effect of measurement noise and temporal resolution on one's ability to quantify the irreversibility of microscopic processes. Specifically, we are interested in estimating the entropy production rate associated with a microscopic process $X(t)$ in a nonequilibrium steady state from an experimental observable $Y(t)$ that may only be known at certain moments of time (for example, when a photon measuring the process is detected) and may be different from $X(t)$ because of noise. The interplay between spatial (or, more generally, state-space) and temporal resolution is important. When the process of interest is slow, averaging over time or super-resolution methods can reduce the effect of noise, but these approaches no longer work for fast dynamics.\cite{PresseRevModPhys} Moreover, even when the true dynamics is a Markov process, the dynamics degraded by noise may be non-Markovian.\cite{Song2024FRET}

\section{Theory}
\subsection{The dynamics of experimental observables have memory even when true dynamics are Markovian}

Let $X(t)$ be the trajectory of a microscopic quantity of interest as a function of time. Typically, $X(t)$ is not accessible directly; rather, a measurement yields an observable trajectory $Y(t)$ that depends on $X(t)$. For example, in single-molecule tracking experiments, $X$ could be the true spatial location of the molecule and $Y$ is the location reported by a single photon emitted by a fluorescent probe attached to the molecule. Or $X$ could be the microscopic state of a molecule and $Y$ is the color of a photon emitted by the molecule while in that state. There are several reasons why $Y$ does not provide the complete information about $X$:
\begin{itemize}
\item Finite temporal resolution: $Y(t)$ is monitored not continuously but only at certain moments of time (e.g., camera capture times or with photon emission times)
\item Experimental noise: the observed values $Y(t)$ differ from $X(t)$ as a result of, e.g., finite spatial resolution (e.g., diffraction limit in single-molecule tracking) or the fact that a photon color does not directly report on the molecular state (e.g., overlapping emission spectra).
\item Coarse graining: Different values of $X$ may correspond to the same $Y$.
 \end{itemize}
While the effects of coarse graining on thermodynamic inference have been discussed in the literature (see, e.g., \cite{SeifertBook,PhysRevX.11.041047,Massi,Rahav_2007,Blom2024,Gomez-Marin2008}), the effects of noise and limited temporal resolution have received very limited attention.

 In what follows, we assume that the true dynamics is a continuous-time Markov process $X(t)$, where $X$ is a discrete or continuous variable. Let $\mathbf{P}(t)$ be the vector representing the probability distribution of $X$. It evolves according to the equation,
\begin{equation} \label{eq:master}
    \frac{d\mathbf{P}(t)}{dt} = \mathbf{K} \mathbf{P}(t),
\end{equation}
where $\mathbf{K}$ is a rate matrix (generator matrix). The general solution of this equation starting from an initial state $\mathbf{P}(0)$ is of the form,
\begin{equation}
    \mathbf{P}(t) = \mathbf{T}(t) \mathbf{P}(0),
\end{equation}
where
\begin{equation}
    \mathbf{T}(t)=e^{\mathbf{K}t}
\end{equation}
is the transition (propagator) matrix. The steady-state distribution of $X$, $\mathbf{P}_{\rm ss}$, satisfies the condition
\begin{equation} \label{eq:steady_state}
\mathbf{K} \mathbf{P}_{\rm ss}=0
\end{equation}
Let us now assume that the value of $X$ is probed at time intervals $t=0,\tau, 2\tau \dots$.\footnote{In fluorescence experiments, photons arrive at random times, with only the average time $\tau$ known. This case can be considered using the same formalism, as will be seen below} Furthermore, we assume that whenever the value $X$ is probed, the measurement yields a value $Y$ that may or may not be different from $X$. The probability of making an error is given by
\begin{equation} \label{eq:error_matrix_def}
    P(Y|X) \equiv R_{YX},
\end{equation}
which can be conveniently thought of as a ``error matrix'' $\mathbf{R}$. For continuous $X$ and $Y$, $P(Y|X)$ is the point spread function.

Our goal is to estimate the entropy production rate,
\begin{equation}
    \langle \dot{S} \rangle = \sum_{X'\ne X} K_{X'X}P_{\rm ss}(X)\log\left[\frac{K_{X'X}P_{\rm ss}(X)}{K_{XX'}P_{\rm ss}(X')}\right],
\end{equation}
from the dynamics of the observable $Y$. There are two general strategies to solving this problem. The first one is the inference of the kinetic matrix $\mathbf{K}$ from the dynamics of $Y$. The second strategy is to take $Y(t)$ as the ``true'' dynamics and apply thermodynamic inference directly to the observed variable. In what follows, we will discuss and compare both approaches in detail, focusing on when they are applicable and whether they can introduce artifacts: for example, can experimental errors lead to mistaking an equilibrium, time-reversible trajectory for a nonequilibrium one?

We start with the statistical properties of the observed time series
$$
    Y_n \equiv Y(n\tau)
$$
corresponding to the (hidden) true sequence of the system's states
$$
X_n\equiv X(n \tau).
$$
$X_n$ is a Markov process characterized by the transition probabilities
\begin{equation} \label{eq:discrete_propagator}
    P(X_{n+1},\tau|X_n) = (e^{\mathbf{K}\tau})_{X_{n+1},X_n} \equiv \mathbf{T}_{X_{n+1},X_n}(\tau) .
\end{equation}
Note that this discrete process has the same steady state as the original, continuous-time process. Indeed, Eq.~\eqref{eq:steady_state} implies that
\begin{equation}
e^{\mathbf{K}\tau} \mathbf{P}_{\rm ss}=\mathbf{P}_{\rm ss}
\end{equation}
{\it In contrast to $X_n$, the process governing $Y_n$ is no longer Markovian, and thus it cannot be completely characterized in terms of a conditional probability $P(Y_{n+1},\tau|Y_n)$}. To illustrate this observation, suppose that $X_n$ measures the consecutive positions of a diffusing particle and that measurements introduce localization errors $\Delta_n$, which are independent, identically distributed variables. Since both $X_n$ and $\Delta_n$ are Markovian, the two-dimensional process $(X_n,\Delta_n)$ is also Markovian, while the measured observable, $Y_n=X_n+\Delta_n$, can be thought of as a one-dimensional projection of $(X_n,\Delta_n)$. It is well known that such projections commonly lead to memory effects. \cite{Zwanzig}

In principle, the statistical properties of $Y$ can be computed if the underlying dynamics, Eq.~\eqref{eq:master} and the error matrix, Eq.~\eqref{eq:error_matrix_def}, are known. In particular, let $P(Y',\tau;Y,0)$ be the joint probability of observing the value $Y$ at $t=0$ and $Y'$ at $t=\tau$. We have
\begin{equation}
   P(Y',\tau;Y,0)= P(Y',\tau|Y)P_{\rm ss}(Y),
\end{equation}
where $P_{\rm ss}(Y)$ is the steady-state probability of Y. These probabilities can be expressed as
\begin{widetext}
\begin{equation} \label{eq:joint_first_order}
    P(Y',\tau;Y,0) = \sum_{X,X'}P(X',\tau;X,0) P(Y|X) P(Y'|X') = \sum_{X,X'}P(X',\tau|X) P_{\rm ss}(X) P(Y|X) P(Y'|X'),
\end{equation}
\end{widetext}
where the summation should be replaced by integration if $X$ is continuous.
Using
\begin{equation}
    P_{\rm ss}(Y)=\sum_{X}P(Y|X)P_{\rm ss}(X),
\end{equation}
we can recover the effective propagator for $Y$

\begin{widetext}
\begin{equation} \label{eq:eff_propagator}
    P(Y',\tau|Y) = \frac{\sum_{X,X'}P(X',\tau|X) P_{\rm ss}(X) P(Y|X) P(Y'|X')}{\sum_{X}P(Y|X)P_{\rm ss}(X)}.
\end{equation}
\end{widetext}
Note that Eq.~\eqref{eq:joint_first_order} can be written in matrix form as
\begin{equation} \label{eq:apparent_propagator}
    \tilde{\mathbf{\rho}}=\mathbf{R} \mathbf{\rho} \mathbf{R}^T,
\end{equation}
where $\mathbf{\rho}$ is the matrix with its elements given by the joint probabilities $\rho_{X'X} \equiv P(X',\tau;X,0)$ and $\tilde{\mathbf{\rho}}$ is the matrix with $\tilde{\rho}_{Y'Y} \equiv P(Y',\tau;Y,0)$.

Importantly, this effective propagator does not provide a complete description of the non-Markovian time series $Y_n$. Nevertheless, higher order joint probabilities can be calculated similarly. For example,
\begin{widetext}
\begin{equation} \label{eq:joint_higher_order}
P(Y'',2\tau; Y',\tau;Y,0) = \sum_{X,X',X''}P(X'',2\tau;X',\tau;X,0) P(Y''|X'')P(Y|X) P(Y'|X')
\end{equation}
\end{widetext}
\subsection{Strategy 1: Inferring microscopic dynamics from noisy trajectories}
If $\mathbf{R}$ is an invertible matrix then the true propagator associated with $X$ can be recovered in two steps as follows. First, using
\begin{equation} \label{eq:invert_propagator}
    \mathbf{\rho} = \mathbf{R}^{-1}\tilde{\mathbf{\rho}}(\mathbf{R}^T)^{-1}.
\end{equation}
we recover the true matrix of joint probabilities of consecutive values of $X$.\footnote{Note that this is a deconvolution problem, which is often notoriously difficult numerically.} Similarly, we recover the vector of steady-state probabilities for $X$
\begin{equation} \label{eq:invert_steady_state}
    \mathbf{\pi} = \mathbf{R}^{-1}\tilde{\mathbf{\pi}},
\end{equation}
where $\tilde{\mathbf{\pi}}$ is a vector whose components are the steady-state probabilities $P_{\rm ss}(Y)$. We now can compute the propagator matrix (Eq.~\eqref{eq:discrete_propagator}),
\begin{equation} \label{eq:recover_T}
 P(X',\tau|X) = \mathbf{T}_{X',X}(\tau)= \rho_{X'X}/P(X).
\end{equation}
 The second step involves recovering the matrix $\mathbf{K}$ as the matrix logarithm of the discrete propagator (i.e., invert Eq.~\eqref{eq:discrete_propagator}): \cite{Bauer_2025}
\begin{equation} \label{eq:matrix_log}
    \mathbf{K}=-\frac{1}{\tau} \log{\mathbf{T}(\tau)}.
\end{equation}
Key requirements for this approach are that the error matrix {\it is known} (with sufficient accuracy) and {\it is invertible}. Furthermore, the step of taking the matrix log numerically is challenging, especially when the time interval $\tau$ exceeds the inverse of some of the eigenvalues of $\mathbf{K}$, in which case the information about short-time dynamics governed by $\mathbf{K}$ is lost in the observed trajectories.\cite{Bauer_2025}

\subsection{Strategy 2: Estimating entropy production directly from noisy observables}
We now consider the possibility of estimating the entropy production using the experimentally observable trajectory $Y(t)$ directly as a proxy for $X(t)$. In doing so, we also assume that the values of $Y(t)$ are only available at discrete time intervals that are multiples of $\tau$. Since the time evolution of $Y(t)$ is not necessarily Markovian, we consider the $k$-th order Markov estimator

\begin{widetext}
\begin{equation} \label{eq:nonmarkov_estimator}
    \langle \dot{S} \rangle^{k}(\tau)=\frac{1}{k\tau} \sum_{Y_0,\dots, Y_{k}}P(Y_k,Y_{k-1},\dots,Y_0)\log\left[ \frac{P(Y_k,Y_{k-1},\dots,Y_0)}{P(Y_0,Y_{1},\dots,Y_k)}\right ],
\end{equation}
\end{widetext}
where $P(Y_k,Y_{k-1},\dots,Y_0)$ is the probability of observing the sequence of states $Y_0,Y_1,\dots,Y_k$.
This estimator has several important properties:
\begin{enumerate}
    \item If $P(Y|X)= \delta_{YX}$ (i.e., no errors in determining the state $X$) then the sequence of observed states $X_k$ is Markovian, and $\langle \dot{S} \rangle^{k}(\tau)$ is the same for any $k\ge 1$. Thus a first-order Markovian estimate suffices.
    \item In the limit $\tau \to 0$ and $P(Y|X)\to \delta_{YX}$ this estimator will give the exact entropy production rate for any $k\ge 1$, $\langle \dot{S} \rangle^{k}(\tau)\to \langle \dot{S} \rangle$.
    \item In the case $P(Y|X)= \delta_{YX}$ we have the inequality $\langle \dot{S} \rangle^{k}\le \langle \dot{S} \rangle$, which becomes identity when $\tau\to 0$.

PROOF:
Since the discrete-time dynamics of $X$ is Markovian, the estimated entropy production can be written as
\begin{equation} \label{eq:nonmarkov_estimator_k1}
    \langle \dot{S} \rangle^{k}(\tau) = \langle \dot{S} \rangle^{1}(\tau)=\frac{1}{\tau} \sum_{X_0, X_{1}}P(X_1,X_0)\log\left[ \frac{P(X_1,X_0)}{P(X_0,X_1)}\right ].
\end{equation}
    Let $X_{1(0),0(1)}(t)$ denote any microscopic path satisfying the conditions $X(0)=X_{0(1)}$ and $X(\tau)=X_{1(0)}$ and let $P[X_{1,0}(t)]$ be its probability measure. Then,
    $$
    P(X_1,X_0)=\sum_{X_{1,0}(t)} P[X_{1,0}(t)]
    $$
    and, according to the log-sum inequality,
    \begin{widetext}
    $$
    \frac{1}{\tau} \sum_{X_0, X_{1}}P(X_1,X_0)\log\left[ \frac{P(X_1,X_0)}{P(X_0,X_1)}\right ]\le \frac{1}{\tau} \sum_{X_{0,1}(t)}P[X_{1,0}(t)]\log\{ \frac{P[X_{1,0}(t)]}{P[X_{0,1}(t)]}\}.
    $$
    \end{widetext}
    But the expression appearing on the rhs of the above inequality is the exact entropy production rate for the Markovian trajectory $X(t)$.

    \item If the microscopic trajectory $X(t)$ satisfies detailed balance then $\langle \dot{S} \rangle^{k}(\tau)=0$. In other words, experimental errors in measuring $X(t)$ or finite time resolution cannot introduce spurious irreversibility in $Y(t)$. 

PROOF:
Using Eq.~\eqref{eq:joint_first_order}, one finds that, if the detailed balance condition $P(X',\tau;X,0)=P(X,\tau;X',0)$ is satisfied for $X(t)$, it is also satisfied for $Y(t)$, $P(Y',\tau;Y,0)= P(Y,\tau;Y',0)$. In fact, since a similar equation can be written for the joint probability of any sequence $Y_0,\dots,Y_k$ (cf. Eq.~\eqref{eq:joint_higher_order}), this probability, $P(Y_k,Y_{k-1},\dots,Y_0)$, is identical to that of the time-reversed sequence $P(Y_0,Y_{1},\dots,Y_k)$. Therefore Eq.~\eqref{eq:nonmarkov_estimator} is equal to zero.

\item Markov estimate of the entropy production from $Y(t)$ is always a lower bound on that estimated from $X(t)$. More precisely,
\begin{widetext}
\begin{equation} \label{eq:ep_Y_bound}
    \frac{1}{\tau} \sum_{X_0, X_{1}}P(X_1,X_0)\log\left[ \frac{P(X_1,X_0)}{P(X_0,X_1)}\right ]\ge \frac{1}{\tau} \sum_{Y_0, Y_{1}}P(Y_1,Y_0)\log\left[ \frac{P(Y_1,Y_0)}{P(Y_0,Y_1)}\right ]
\end{equation}
\end{widetext}
PROOF:
First, let us prove the following auxiliary statement, which is a form of the data processing inequality. Let $x$ be a random vector with a distribution $p(x)$ and let $y$ be a vector with its distribution given by
$$
\tilde{p}(y)=\sum_x K(y|x)p(x),
$$
where $K(y|x)$ is a stochastic matrix (i.e., a matrix of transition probabilities from $x$ to $y$). Similarly, let $q(x)$ be another distribution and
$$
\tilde{q}(y)=\sum_x K(y|x)q(x),
$$
then
\begin{equation} \label{eq:data_processing}
\sum_x p(x) \log \frac{p(x)}{q(x)} \ge \sum_y\tilde{p}(y) \log \frac{\tilde{p}(y)}{\tilde{q}(y)}.
\end{equation}
To prove this we start with the log sum inequality written as
$$
\sum_{i}a_i \log\frac{a_i}{b_i}\ge (\sum_{i}a_i) \log\frac{\sum_ia_i}{\sum_i b_i}
$$
For some value $y$ let $a_x=K(y|x)p(x)$, $b_x=K(y|x)q(x)$. Substituting this into the log sum inequality, we get
\begin{widetext}
$$
\sum_x K(y|x)p(x) \log\frac{p(x)}{q(x)} \ge \{\sum_x K(y|x)p(x)\} \log \frac{\tilde{p}(y)}{\tilde{q}(y)} =\tilde{p}(y) \log \frac{\tilde{p}(y)}{\tilde{q}(y)}.
$$
\end{widetext}
Now summing over $y$ and using the fact that $\sum_y K(y|x)=1$ we obtain Eq.~\eqref{eq:data_processing}.

Eq.~\eqref{eq:ep_Y_bound} now follows from Eq.~\eqref{eq:data_processing} if we set $x=(X_1,X_0)$, $y=(Y_1,Y_0)$, $q(x)\equiv Q(X_2,X_1)=P(X_1,X_2)$, $\tilde{q}(y)\equiv Q(Y_1,Y_2)=P(Y_2,Y_1)$. In other words, $p$ is the distribution of a pair of observations of $X$ and $q$ is the distribution of the time reversed observations.

\item The above argument can be extended to higher-order entropy production estimates. The entropy production estimated from sequences $Y_0,Y_1\dots Y_k$ will always be lower than or equal to that from sequences $X_0,X_1\dots X_k$. But the latter does not depend on $k$ since $X$ is Markovian. We conclude that the entropy production estimated from ${Y_i}$ to arbitrary Markov order is always a lower bound on the true entropy production.

\end{enumerate}

\section{Examples}

\subsection{Case study: Using strategy 1 to estimate the entropy production for continuous-time random walks on a network}
\begin{figure}
    \centering
    \includegraphics[width=0.9\linewidth]{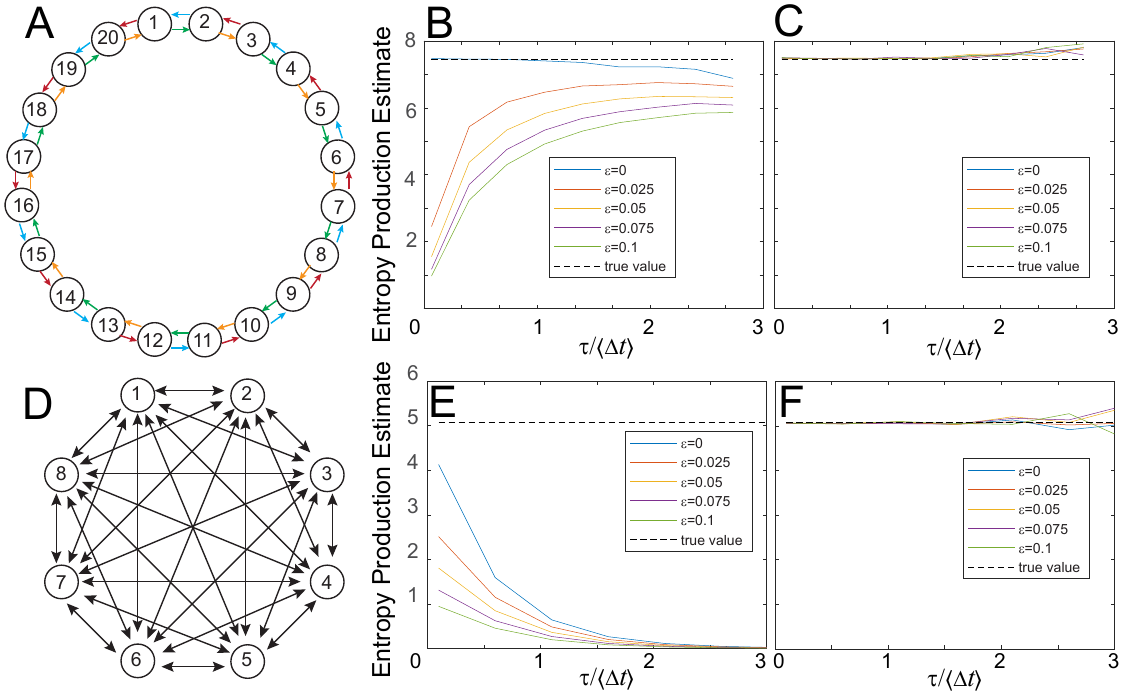}
    \caption{Entropy production estimates calculations. (A-C) 20-node ring with 4 distinct rates (shown as arrows of different colors); (B) Direct first-order Markov estimation of entropy production from the noisy trajectories. (C) Entropy production calculated using the recovered transition-rate matrix computed with Eq.~\eqref{eq:matrix_log} (D-F) 8-node fully connected network. (E) Direct first-order Markov estimation of entropy production from the noisy trajectories. (F) Entropy production calculated using the recovered transition-rate matrix computed with Eq.~\eqref{eq:matrix_log}. Different colors show different noise levels as indicated in the legends.}
    \label{fig:network_estimates}
\end{figure}

We start with the two networks' topologies shown in Fig.~\ref{fig:network_estimates}. For each we follow the same procedure

\begin{enumerate}
    \item Generate noiseless trajectory with $2\times 10^7$ transitions using the Gillespie algorithm
    \item Sample the trajectory at time intervals $\tau$
    \item Introduce measurement noise every time step
    \item Compute transition probabilities from the uncorrected trajectories and estimate entropy production with the first-order Markov estimator
    \item Compute the corrected entropy production using the recovered transition-rate matrix computed with Eq.~\eqref{eq:matrix_log}.
\end{enumerate}

First, we consider a walker on a ring, i.e., a simple 1-D chemical reaction network with periodic boundary conditions (Fig.~\ref{fig:network_estimates}A),
$1\leftrightarrow 2\leftrightarrow\dots \leftrightarrow N \leftrightarrow 1$. This network can describe, for instance, the sequence of rotational states of a rotary molecular motor such as F1FO ATPase, bacterial flagella motor,\cite{oster2003rotary} or a sequence of biochemical states for translational motor proteins.\cite{TolyaBook} We denote the forward rate for the $i\to i+1$ reaction as $k_i\equiv k_{i\to i+1}$, and the backward rate for the $i\to i-1$ reaction as $k_{-i}\equiv k_{i\to i-1}$. Periodic boundary conditions imply that indices are defined $\mod N$, i.e., $N\equiv 0$ and $1\equiv N+1$.

The walker's state (position) is measured at time intervals $\tau$, with the correct position $i$ recovered with a probability $1-2\epsilon$, and with two possible incorrect positions at the nearest neighbor locations $i\pm 1$ registered with probability $\epsilon$ each. Thus, the error matrix is $N \times N$ and it has $1-2\epsilon$ on the diagonal, and $\epsilon$ above the diagonal, below the diagonal, and at $(1,N)$/$(N,1)$ positions
\begin{equation} \label{eq:error_matrix_ring}
    \mathbf{R}=\begin{pmatrix}
1-2\epsilon & \epsilon & 0& \cdots &0 &\epsilon \\
\epsilon & 1-2\epsilon & \epsilon&0 &\cdots& 0 \\
0& \epsilon & 1-2\epsilon & \epsilon &\cdots& 0&  \\
\vdots& \vdots & \vdots & \vdots &\vdots& \vdots&  \\
0 &\cdots&0 &\epsilon & 1-2\epsilon&0\\
\epsilon &0&\cdots&0 &\epsilon & 1-2\epsilon
\end{pmatrix}.
\end{equation}

For numerical simulations, we have chosen $N=20$ with $k_1= 4.1668$, $k_{-1}=0.7978$, $k_2=7.4015$, and $k_{-2}=1.2722$. For the remaining rates, we assumed a 2-step periodicity: $k_{\pm(i+2)}=k_{\pm i}$. The above-described procedure is repeated with varying $\tau$ values ranging
from $0.01 \langle\Delta t \rangle$ to $3\langle \Delta t \rangle$ ( $\langle \Delta t \rangle$ is the mean transition time of the noiseless trajectory) and with a noise term with magnitude $\epsilon$ ranging from $0$ to $0.1$.

Figure \ref{fig:network_estimates}B,C compares the entropy production estimated directly from the noisy signal $Y(k\tau)$ using the Markovian estimator (Eq.~\eqref{eq:nonmarkov_estimator} with $k=1$) with the entropy production computed using the matrix $\mathbf{K}$ estimate recovered from Eqs.~\eqref{eq:invert_propagator}-\eqref{eq:matrix_log} and with the exact entropy production using the true rate matrix $\mathbf{K}$.

As expected, without noise and for sufficiently small $\tau$, the true entropy production is readily recovered directly from $Y(t)=X(t)$, Fig.~\ref{fig:network_estimates}B. In the presence of noise, however, using $Y(k \tau)$ with small time intervals $\tau$ leads to drastic underestimation of the entropy production. This amplification of the noise effect at short time intervals is easy to understand. For example, if $X(t)$ only moves clockwise every time moment when its location is measured, the measured trajectory $Y(t)$ will instead contain clockwise and counterclockwise steps of length $\pm 1$ caused by the measurement noise. Because these spurious steps have finite length in our model, measuring the walker locations at longer time intervals $\tau$ will reveal the overall clockwise motion, reduce the impact of the localization error, and thus improve the entropy production estimate. This is similar to using wider-spaced milestones in the milestoning method\cite{Blom2024} to reduce the effects of non-Markovianity of the observed time series. As will be seen in the example studied in Section IIIB, sometimes increasing the measurement time interval is all that is needed to estimate the entropy production with any desired accuracy. Here, however, $\tau$ cannot be increased indefinitely: when it is too large, then a full rotation along the circle, happening while the walker is unobserved, could be missed. This limits the accuracy of the entropy production estimates that can be obtained with the observable time series $Y(t)$, as observed in Fig.~\ref{fig:network_estimates}B.

In contrast, when Eqs.~\eqref{eq:invert_propagator}-\eqref{eq:matrix_log} are used to recover the ``microscopic rate matrix'' $\mathbf{K}$, the entropy production is recovered to within $1\%$ for all values of $\epsilon$ considered, as long as $\tau$ is less than $\approx 1.5 \langle\Delta t \rangle$, where $\Delta t$ is the mean time between transitions ( Fig.~\ref{fig:network_estimates}C ). For $\tau>\approx 2.5 \langle\Delta t \rangle$, numerical issues with computing the matrix logarithm led to incorrect estimates. 

For a second, more general case study, we consider a fully connected network with $N=8$ states, and $N(N-1)=56$ possible transitions. We again assume the error matrix in the form of $\mathbf{R}$ from Eq.~\eqref{eq:error_matrix_ring}. We generate all the transition rates from a log-normal distribution ($\log$-mean 0, $\log$-variance 1; resulting matrix is given in Supporting information) and follow the same procedure to generate trajectories for different $\tau$ and $\epsilon$ values.
The results shown in Fig.~\ref{fig:network_estimates}E indicate that even without measurement noise $\epsilon=0$ and at the smallest $\tau$ used, $\tau \sim 0.1\langle\Delta t \rangle$, the uncorrected estimate of entropy production is at least $20\%$ smaller than the true value. An increase in $\epsilon$ and $\tau$ makes the apparent (uncorrected) entropy production quickly decay to $0$. Notably, in contrast to the previous example, in the fully connected network of Fig.~\ref{fig:network_estimates}D, increasing the sampling time interval $\tau$ further degrades rather than improves thermodynamic inference. In contrast, our correction procedure allowed us to precisely recover the entropy production value for all $\epsilon$ considered as long as $\tau$ is less than $\approx 1.5 \langle\Delta t \rangle$ (Fig.~\ref{fig:network_estimates}F). Further increases of $\tau$ again led to numerical issues with computing the matrix logarithm and incorrect estimates.

\begin{figure}
    \centering
    \includegraphics[width=0.95\linewidth]{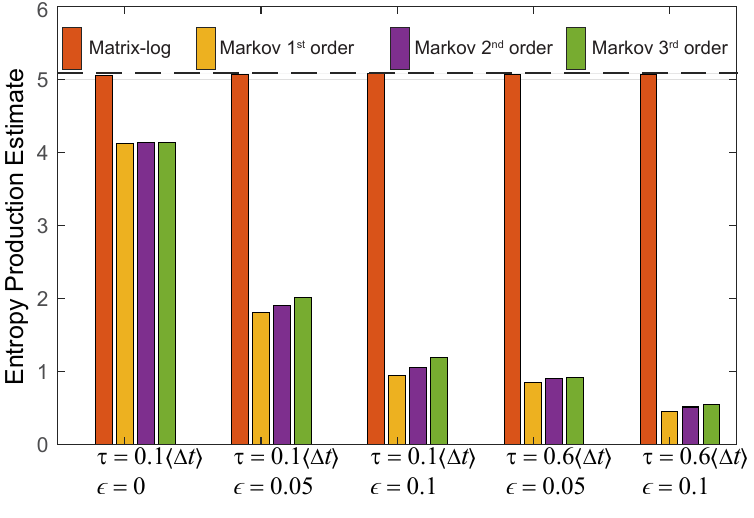}
    \caption{Non-Markov entropy production estimates, Eq.~\eqref{eq:nonmarkov_estimator}, applied to the 8-state system and compared to the exact entropy production as well as to the corrected entropy production estimates.}
    \label{fig:nonmarkov_bar}
\end{figure}

Although the two-step correction method works quite well in this case study, it relies on accurate knowledge of the error matrix and its invertibility. Applying an entropy production estimator directly to the experimental data $Y$ is attractive, as it does not rely on such assumptions. As observed in Fig.~\ref{fig:network_estimates}, a Markov estimator significantly underestimates the entropy production. As $Y(t)$ is non-Markovian, a natural question to ask is whether using a higher-order estimator improves this estimate significantly. While this issue will be explored in more detail in Section IIIC for a simpler model, here we apply a non-Markov estimator, Eq.~\eqref{eq:nonmarkov_estimator} to the same 8-state system, Fig.~\ref{fig:nonmarkov_bar}. It should be noted that estimating Eq.~\eqref{eq:nonmarkov_estimator} at high Markov orders $k$ is not only computationally expensive but also requires large amounts of data to estimate the joint probabilities that appear in this equation. For this reason, using the same trajectories as those used to generate Fig.~\ref{fig:network_estimates} we could only obtain such estimates for $k\le 3$.

Depending on the time step $\tau$, we used two versions of the method. For short time interval $\tau$ we used the stroboscopic time series $Y_k \equiv Y(k \tau)$ in the non-Markov estimator of Eq.~\eqref{eq:nonmarkov_estimator}. Without measurement noise, this generates the ``uncorrected'' Markov estimate (i.e., without application of Eq.~\eqref{eq:matrix_log}) , as indeed is observed in Fig.~\ref{fig:nonmarkov_bar}. For longer time intervals, comparable to the mean time between consecutive transitions, we first computed, from $Y_k$ a sequence $Y'_n$ of {\it distinct} states visited by the system\cite{Blom2024,two-track-walk}. For example, if $Y_k$ contains a sequence $\ldots 11122345\ldots$, the corresponding sequence $Y'_n$ is $\ldots 12345\ldots$. The non-Markov estimator of Eq.~\eqref{eq:nonmarkov_estimator} was then applied to $Y'_n$ (with $\tau$ being replaced with the mean time between transitions between distinct states in the $Y'_n$ sequence).

A key observation from Fig.~\ref{fig:nonmarkov_bar} is that including non-Markov effects to modest Markov order (dictated by statistics, CPU, and memory limitations) produces only a small improvement to the entropy production estimates, especially when compared to the results based on inferring the true propagator matrix for the underlying trajectory $X(t)$ presented in Fig.~\ref{fig:network_estimates}.

\subsection{Case study: Estimating entropy production for diffusion in the presence of a constant force}

Consider a Brownian particle subjected to a constant force $F$. Its Green's function is the solution of the diffusion equation with a drift, given by:
\begin{equation} \label{eq:green_diffusion}
    P(X',\tau|X)=\frac{1}{\sqrt{4\pi D \tau}} \exp{-\frac{(X-X'-\beta F D \tau)^2}{4 D \tau}},
\end{equation}
where $D$ is the particle's diffusivity and $\beta=(k_B T)^{-1}$ is the inverse thermal energy. Let the point spread function $P(Y|X)$ be a Gaussian,

\begin{equation} \label{eq:psf}
    p(Y|X)=\frac{1}{\sqrt{2\pi \sigma^2}} \exp{-\frac{(Y-X)^2}{2 \sigma^2}},
\end{equation}
Using Eq.~\eqref{eq:eff_propagator}, we obtain the effective propagator,
\begin{equation}
P(Y',\tau|Y)=\frac{1}{\sqrt{4\pi (D \tau+\sigma^2)}} \exp{-\frac{(Y-Y'-\beta F D \tau)^2}{4 (D \tau+\sigma^2)}}.
\end{equation}
In this particular case, but not generally, the exact entropy production can be recovered from this propagator using our second strategy. Indeed, treating $Y(t)$ as the true dynamics, we estimate the associated entropy production as:
\begin{equation} \label{eq:ep_force}
    \langle \dot{S} \rangle^{(1)}=\frac{1}{\tau}\int_{-\infty}^{\infty}dY'P(Y',\tau|Y) \ln{\frac{P(Y',\tau|Y)}{P(Y,\tau|Y')}} = \frac{v^2\tau}{\sigma^2+D\tau},
\end{equation}
where
\begin{equation} \label{eq:drift_velocity}
    v = F\beta D
\end{equation}
is the particle's mean drift velocity. If we take measurements infrequently, $\tau\gg \sigma^2/D$, then the {\it exact} entropy production is recovered.
\begin{equation}
    \langle \dot{S} \rangle^{(1)} = \langle \dot{S} \rangle=\frac{v^2}{D}.
\end{equation}
Here, the lucky break comes because of the translational invariance of the dynamics, similarly to the case of random walks on a ring discussed in the previous Section. This is also a manifestation of the general idea that, sometimes, coarse-graining by discarding fine-scale information may improve thermodynamic inference.\cite{Blom2024} However, this approach has obvious limitations, as exemplified by the second example of the previous section, where information about local currents contributing to the entropy production is discarded in a coarse-grained view of the dynamics.

\subsection{Case study: non-Markov estimator of the entropy production for a discrete random walk with 3 states}
To study to what extent the effects of the noise-induced non-Markovianity can be mitigated by using non-Markov estimators, here we consider the toy model of a random walker that can occupy 3 sites (numbered 1, 2, and 3) forming a ring. The walker steps in the clockwise direction with a probability $p$ and counterclockwise with probability $q=1-p$. At every step the location of the walker is measured, with the correct position being measured with a probability $1-2\epsilon$ and two possible wrong positions registered with probabilities equal to $\epsilon$. The ``error'' matrix therefore has the same form as in Eq.~\eqref{eq:error_matrix_ring},
\begin{equation} \label{eq:error_matrix_3state}
    \mathbf{R}=\begin{pmatrix}
1-2\epsilon & \epsilon & \epsilon \\
\epsilon & 1-2\epsilon & \epsilon \\
\epsilon & \epsilon & 1-2\epsilon
\end{pmatrix}
\end{equation}
and the matrix of joint probabilities $\rho$ is
\begin{equation}
    \mathbf{\rho}=\frac{1}{3}\begin{pmatrix}
0 & 1-p & p \\
p & 0 & 1-p \\
1-p & p & 0
\end{pmatrix}
\end{equation}
The apparent propagator describing the observed process is given by Eq.~\eqref{eq:apparent_propagator}. Using this propagator we can estimate the apparent entropy production, which is given by
\begin{equation} \label{eq:ep_first_order_3state}
    \langle \dot{S} \rangle^{(1)}(\epsilon,p)=(1-2p)(1-3\epsilon)^2 \ln{\frac{1-p(1-3\epsilon)^2-4\epsilon+6\epsilon^2}{p(1-3\epsilon)^2+(2-3\epsilon)\epsilon}}.
\end{equation}
Notably, this expression is nonzero for $0<\epsilon<1/2$ except the point $\epsilon=1/3$, where it vanishes. It is easy to see why the apparent entropy production vanishes in this case: all the entries in the matrix of Eq.~\eqref{eq:error_matrix_3state} are identical and equal to $1/3$, which means that a measurement simply replaces $X$ by $Y$ drawn with identical probabilities thus removing any information about the original value of $X$ and thus about the original process. One can also show that the value predicted by Eq.~\eqref{eq:ep_first_order_3state} is always less than the true entropy production equal to $\langle \dot{S} \rangle=(2p-1)\ln{\frac{p}{1-p}}$.

\begin{figure}
    \centering
    \includegraphics[width=0.75\linewidth]{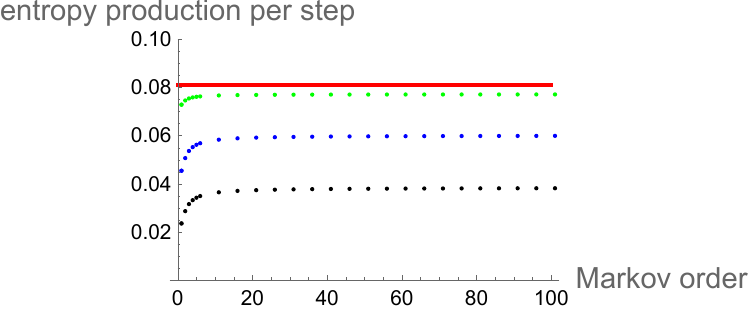}
    \caption{Entropy production estimate, Eq.~\eqref{eq:nonmarkov_estimator}, as a function of the Markov order $k$, for the toy 3-state discrete random walk with $p=0.6$ and with $\epsilon=0.01$ (green), $\epsilon=0.05$ (blue), and $\epsilon=0.1$ (black). The horizontal line shows the exact entropy production in this case}
    \label{fig:three_state_markov_order}
\end{figure}

As shown in Fig.~\ref{fig:three_state_markov_order} using a non-Markov estimator, Eq.~\eqref{eq:nonmarkov_estimator} improves on the first-order estimate; in the limit $k\to\infty$ the value $\langle \dot{S} \rangle^{(k)}(\epsilon,p)$ approaches a limit that is still below the true value.
We note that high order estimates are expensive computationally.\cite{roldan2012entropy,Roldan2010} Here, we have been able to get high order estimates by using a hybrid approach (see, e.g., \cite{two-track-walk}) which takes advantage of the knowledge of the underlying true dynamics $X$; as such this method is not useful in practice.

We also note that using Eqs.~\eqref{eq:invert_propagator} we have been able to trivially recover the true propagator of the system and thus recover the full entropy production with an accuracy of a few percent from random walks of $10^5$ steps (data not shown).

\section{Entropy production from photon emission}
This Section focuses on thermodynamic inference for a specific experimental setup using single-molecule FRET (fluorescence resonance energy transfer). There have been considerable efforts in using this approach to study microscopic dynamics of biomolecular machines (see Ref.\cite{HARAN2026103242} for a recent review), and several recent papers focused specifically on inferring the arrow of time from such measurements\cite{Hugel_HMM, Hugel2014, Song2024FRET}. FRET differentiates among the states of a system on the basis of the colors of photons emitted by a fluorescent probe attached to the molecule. Here we consider a model of a 3-state rotary molecular motor\cite{kolomeisky2013motor} that makes clockwise and counter-clockwise transitions with rates $u$ and $w$, a model that is similar to that of Section IIIA and is a continuous-time version of the 3-state random walk considered in Section IIIC. The rate matrix of this system is given by
\begin{equation}
    \label{eq:three_state_generator}
    \mathbf{K}=\begin{pmatrix}
-u-w & w & u \\
u & -u-w & w \\
w & u & -u-w
\end{pmatrix}
\end{equation}
The state of the motor is deduced from the color of the photons emitted by the system. Unlike the stroboscopic measurements considered in Section IIIC, the photon arrival times are also random variables. We consider the 3 color FRET, where the colors $Y=1,2,3$ are used to report on the state of the motor, $X=1,2,3$. Ideally, we wish to have one-to-one correspondence between the colors and the states, $X=Y$, but in any realistic FRET experiment a mix of colors is detected for any given molecular state $X$. As before, the probability $P(Y|X)$ that a photon of color $Y$ is detected while the motor is in a state $X$ is given by the elements of the error matrix $\bf{R}$. We further assume that the photon arrival times are governed by Poisson statistics, with a photoemission rate of $\mu$.

We now show that it is possible, in principle, to recover the exact rate matrix $\bf{K}$ from the sequence of photon colors, provided that the error matrix $\bf{R}$ is invertible. Indeed, consider first the sequence of states $\{X_i\}$ occupied by the system at the moments of time when the photons are emitted. It is a discrete-time Markov process governed by the propagator matrix
\begin{equation}
\begin{split}
    P(X_{i+1}|X_i)\equiv \mathbf{T}_{X_{i+1},X_{i}}\\
    =\mu \int_{0}^{\infty}e^{-\mu t}(e^{\mathbf{K}t})_{X_{i+1},X_i}dt\\
    =\mu (\mathbf{K}-\mu\mathbf{1})^{-1}_{X_{i+1},X_i},
    \end{split}
\end{equation}
where $\mathbf{1}$ is the $3\times 3$ identity matrix. Therefore, if the propagator matrix is known then we can recover the matrix $\mathbf{K}$ via matrix inversion of $\mathbf{T}$,
\begin{equation} \label{eq:K_from_T}
    \mathbf{K}=\mu^{-1}\mathbf{T}^{-1}+\mu \mathbf{1}.
\end{equation}
However, the sequence of observed photon colors is generally non-Markovian. Nevertheless, if the transition probabilities for this process are known (cf. Eq.~\eqref{eq:eff_propagator}),
\begin{equation} \label{eq:eff_propagator_photons}
    P(Y'|Y) = \frac{\sum_{X,X'}P(X'|X) P_{\rm ss}(X) P(Y|X) P(Y'|X')}{\sum_{X}P(Y|X)P_{\rm ss}(X)},
\end{equation}
and if the error matrix is known and is invertible, the underlying propagator matrix $\mathbf{T}$ can be recovered using the method of Strategy~1 (Eqs.~\eqref{eq:invert_propagator}-\eqref{eq:recover_T}), where the joint probability $P(Y',\tau; Y,0)$ of making observations $Y$ and $Y'$ separated by a time interval $\tau$ should be replaced by the joint probability $P(Y',Y)$ of successive photon colors being $Y$ and $Y'$, and where the respective conditional probabilities are similarly modified. That is, recovering entropy production from the statistics of photon sequences is analogous to the method of Strategy~1 except that evaluation of a matrix logarithm, Eq.~\eqref{eq:matrix_log}, must be replaced by matrix inversion, Eq.~\eqref{eq:K_from_T}.

The above method requires an accurate estimate of the error matrix, which is not always possible. Moreover, in FRET experiments the number of photon colors is often smaller than the number of states, and so the error matrix is not invertible. In this case, one has to resort to estimating the entropy production directly from the statistics of the observed photon colors $Y$. In particular, the Markov entropy production estimator is given by (cf. Eq.~\eqref{eq:nonmarkov_estimator})
\begin{equation} \label{eq:markov_estimator_photons}
    \langle \dot{S} \rangle^{1}=\mu \sum_{Y_0,Y_{1}}P(Y_1,Y_0)\log\left[ \frac{P(Y_1,Y_0)}{P(Y_0,Y_{1})}\right ].
\end{equation}
which is obtained from the first-order Markov estimator for stroboscopic measurements by replacing $1/\tau$ with the photoemission rate $\mu$.

To gain further insight into this estimator, it is useful to recast the same problem as a continuous-time process. Let $X(t) = 1, 2, 3$ be the state of the motor at time $t$ and $Y(t) = 1, 2, 3$ the color of the most recently emitted photon. Between emissions $Y(t)$ is constant, and at each emission it jumps to the newly detected color; sampling at the emission times $t_n$ recovers the discrete sequences of true states and photon colors, $X_n = X(t_n)$ and $Y_n = Y(t_n)$. The joint process $(X(t), Y(t))$ is Markovian, while the observed $Y(t)$ on its own is not. In this picture the state $X = i$ emits the photon $i$ with rate $k$ and produces the other two photons with rate $\delta k$, corresponding to a total photon emission rate of $\mu=k(1+2\delta)$ and to an error matrix of the form of Eq.~\eqref{eq:error_matrix_3state} with per-emission error probability $\epsilon = \delta/(1+2\delta)$. The $\epsilon = 0$ limit corresponds to perfect color reporting, and $\epsilon = 1/3$ is the destructive limit at which all three photon emission rates are equal, so the photon color carries no information about $X$; when $\epsilon$ falls between $1/3$ and $1/2$ the model is somewhat perverse, in that a photon is more likely than not to misrepresent the true state. Note that there is ``stroboscopic'' noise from $X$ advancing before a photon is ever observed, which could appear as a backwards $X$ step when in reality it stemmed from an unobserved $X$ step.

The dynamics in the $(X,Y)$ space is described by a 9-state continuous-time master equation with the states defined by their values of $X$ and $Y$. Let $P(X, Y)$ denote the probability of the state and we choose the basis such that $\mathbf{P} = (P(1,1), P(1,2), P(1,3), P(2,1), \dots, P(3,3))^{T}$. The equation of motion is
\begin{equation} \label{eq:nine_state_generator}
\frac{\partial \mathbf{P}}{\partial t} =
\begin{pmatrix}
 \alpha & k & k & w & 0 & 0 & u & 0 & 0 \\
 k\delta & \beta & k\delta & 0 & w & 0 & 0 & u & 0 \\
 k\delta & k\delta & \beta & 0 & 0 & w & 0 & 0 & u \\
 u & 0 & 0 & \beta & k\delta & k\delta & w & 0 & 0 \\
 0 & u & 0 & k & \alpha & k & 0 & w & 0 \\
 0 & 0 & u & k\delta & k\delta & \beta & 0 & 0 & w \\
 w & 0 & 0 & u & 0 & 0 & \beta & k\delta & k\delta \\
 0 & w & 0 & 0 & u & 0 & k\delta & \beta & k\delta \\
 0 & 0 & w & 0 & 0 & u & k & k & \alpha
\end{pmatrix}
\mathbf{P},
\end{equation}
where $\alpha = -u-w-2k\delta$ and $\beta = -u-w-k(1+\delta)$ are the escape rates for states with $X = Y$ and $X \neq Y$, respectively. The steady-state distribution can be found explicitly to be $\mathbf{P}_{\rm ss} = (a, b, c, c, a, b, b, c, a)^{T},$ where
\begin{align}
a &= \frac{
\mu^2 + 2\mu(u + w) + u^2 + uw + w^2 - \epsilon\mu(2\mu + 3u + 3w)
}{3\mathcal{Z}} \\
b &= \frac{
u^2 + uw + w^2 + \mu w + \epsilon\mu(\mu + 3u)
}{3\mathcal{Z}} \\
c &= \frac{
u^2 + uw + w^2 + \mu u + \epsilon\mu(\mu + 3w)
}{3\mathcal{Z}},
\end{align}
The numerator,
\begin{equation} \label{eq:Z_def}
    \mathcal{Z} = 3(u^2 + uw + w^2) + 3\mu(u+w) + \mu^2
\end{equation}
ensures the normalization condition, $3(a + b + c) = 1$.

In an experiment, only transitions in $Y$ are visible; the $X$ transitions are hidden.
The explicit form of $\mathbf{P}_{\rm ss}$ can be traced over $X$ to give photon color sequence marginals from which the first-order Markov entropy production estimator (cf.\ Eq.~\eqref{eq:markov_estimator_photons}) is computed:
\begin{equation} \label{eq:markov_ep_Y}
    \langle \dot{S} \rangle^{1} = \mu \sum_{Y_0, Y_1} P(Y_1, Y_0) \log\left[\frac{P(Y_1, Y_0)}{P(Y_0, Y_1)}\right].
\end{equation}
The transition probabilities $P(Y_1,Y_0)$ are obtained from the steady-state fluxes of the $(X,Y)$ master equation. The result is a long expression that is equivalent to Eq. 19 in \cite{Song2024FRET} with the mapping between notations: $u=k_F$, $w=k_B$, $\epsilon = p /2$.

The observed photon color sequence induces an effective current $j_Y$, defined as the net rate of clockwise transitions in $Y$-space. Using the steady-state fluxes of the 9-state master equation, this mean current can be computed explicitly. An efficient way to perform the calculation utilizes the Koza method (Appendix~\ref{app:koza}) to obtain
\begin{equation} \label{eq:mean_current_Y}
    \langle j_Y \rangle = \frac{\mu^2(u - w)(1 - 3\epsilon)^2}{\mathcal{Z}}.
\end{equation}
Recall that the denominator is a polynomial depending only on the rates $u, w, \mu$ alone---it has no dependence on the noise $\epsilon$. The noise enters $\langle j_Y\rangle$ only through the prefactor $(1-3\epsilon)^2$, which vanishes at $\epsilon = 1/3$, consistent with the destruction of color information at that point.

  The variance rate (diffusivity) of $j_Y$ is given by the second cumulant of the scaled cumulant generating function. Using the cyclic symmetry to reduce the 9$\times$9 tilted generator to a $3\times 3$ matrix and applying the Koza method~\cite{Koza1999,Koza2002}, one obtains
\begin{equation} \label{eq:variance_rate}
\langle \delta j_Y^2 \rangle = \frac{\mu (1 - 2 \epsilon)\mathcal{N}}{\mathcal{Z}^3},
\end{equation}
where $\mathcal{N}$ is a polynomial in $u, w, \mu, \epsilon$ given explicitly in Appendix~\ref{app:koza}. The prefactor $\mu(1-2\epsilon)$ is the rate of matching-color photon emission.

Given the mean and variance, one can estimate a bound on the entropy production in a manner quite distinct from the Markov estimator Eq.~\eqref{eq:markov_ep_Y}.
This alternative estimator, which can also be applied directly to the observed photon sequences, $\langle \dot{S}\rangle_{\rm TUR}$ follows from the thermodynamic uncertainty relation (TUR)\cite{BaratoSeifert2015,PhysRevLett.116.120601}:
\begin{equation} \label{eq:tur_def}
  \langle \dot{S} \rangle \geq \frac{2\langle j_Y \rangle^2}{\langle \delta j_Y^2 \rangle} \equiv \langle \dot{S} \rangle_{\mathrm{TUR}}.
\end{equation}
Plugging in expressions for the current fluctuations, we have
\begin{equation} \label{eq:TUR_photon}
\langle \dot{S}\rangle_{\rm TUR} = \frac{2\mu^3(u-w)^2(1-3\epsilon)^4\,\mathcal{Z}}{(1-2\epsilon)\mathcal{N}}.
\end{equation}

\begin{figure}
\centering
\includegraphics[width=0.45\textwidth]{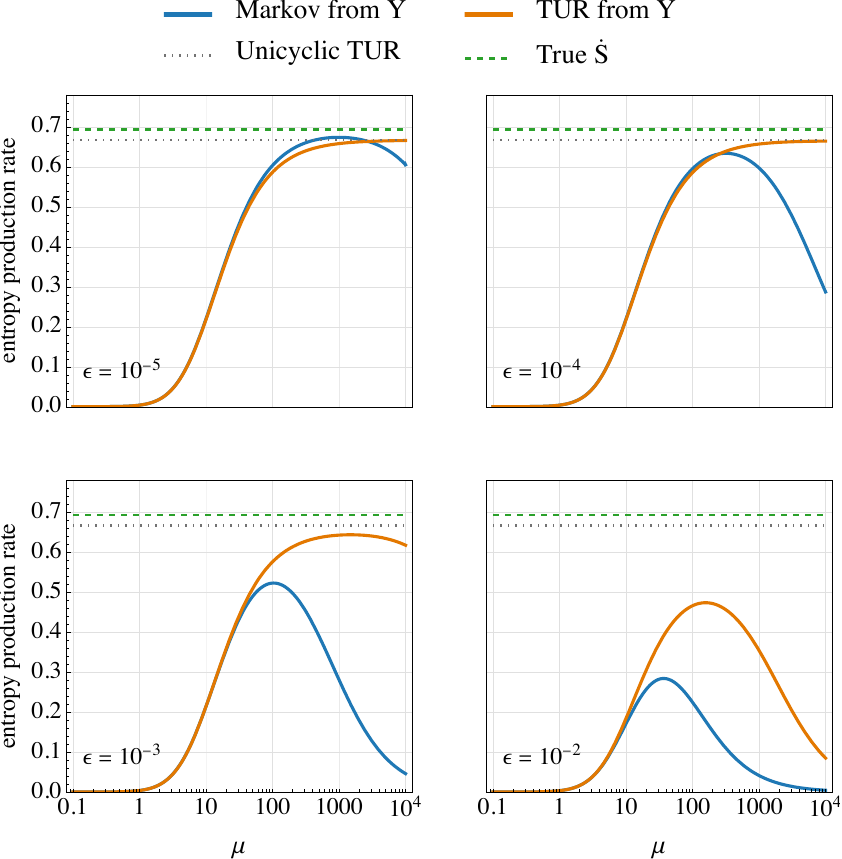}
\caption{Entropy production estimated from the photon color sequence as a function of the total photon arrival rate $\mu$, for $u = 2$, $w = 1$, using the Markov estimator (Eq.~\eqref{eq:markov_ep_Y}) and the TUR estimator (Eq.~\eqref{eq:tur_def}). Panels correspond to increasing values of the per-emission error probability $\epsilon$. The dashed green line is the true entropy production $\langle \dot{S}\rangle = \ln 2 \approx 0.693$, and the dotted gray line is the unicyclic TUR bound $2(u-w)^2/(u+w) \approx 0.667$ of Eq.~\eqref{eq:tur_nested}---the TUR estimate that a perfect observation of the motor's own current would yield. The gap between this bound and $\langle \dot{S}\rangle$ is the intrinsic looseness of the TUR and is independent of the measurement; the gap between the TUR estimate and this bound is the measurement penalty. At fixed $\epsilon>0$ both estimators are non-monotonic in $\mu$, rising to an optimum and then decreasing as the wrong-color noise accumulates and the photon sequence becomes strongly non-Markovian.}
\label{fig:photon_ep}
\end{figure}

The resulting estimates of the entropy production rate lower bound, as a function of the photon emission rate, are shown in Fig.~\ref{fig:photon_ep}. As expected, the quality of the estimates becomes increasingly poorer as the error probability increases. The Markov estimator exhibits a maximum at an ``optimal'' value of the emission rate. The initial increase in the estimated entropy production with the emission rate can be understood as a result of the state of the molecule being ``hidden'' during the times between the photons. The decrease in the estimated entropy production at higher values of the photon emission rate is, on the other hand, a result of the strongly non-Markovian character of photon sequences when observed with an increasing time resolution, as the inter-photon times become shorter \cite{Song2024FRET}, which is not captured by a Markov estimator. As shown in Ref.\cite{Song2024FRET}, higher-order Markov estimators improve the estimate, but often call for unrealistically long trajectories and significant computational expense.

The TUR estimator $\langle \dot{S}\rangle_{\rm TUR}$ has a transparent structure that the Markov estimator lacks: the bound it provides decomposes into two nested inequalities,
\begin{equation} \label{eq:tur_nested}
\frac{2\langle j_Y \rangle^2}{\langle \delta j_Y^2 \rangle} \;\leq\; \frac{2(u-w)^2}{u+w} \;\leq\; (u-w)\ln\frac{u}{w} = \langle \dot{S}\rangle.
\end{equation}
The second inequality is the standard unicyclic TUR for the bare 3-state motor. The first is an information-processing bound: observing the motor only through the noisy, finite-rate photon channel can only degrade the signal-to-noise ratio $2\langle j_Y \rangle^2/\langle \delta j_Y^2 \rangle$ relative to that of the motor's own current, $2(u-w)^2/(u+w)$. We prove the first inequality for this model in Appendix~\ref{app:TUR_proof}.

This nesting is what makes the TUR estimator interpretable. The intermediate quantity $2(u-w)^2/(u+w)$ is the \emph{unicyclic TUR bound} of the bare motor---the value $2\langle j_X\rangle^2/\langle \delta j_X^2\rangle$ that the TUR returns from the motor's own cycle current, with $\langle j_X\rangle = u-w$ and $\langle \delta j_X^2\rangle = u+w$. It depends only on the motor rates $u$ and $w$, with no dependence on the measurement parameters $\mu$ and $\epsilon$. Its shortfall from the true rate $\langle \dot{S}\rangle$ is therefore purely thermodynamic---the intrinsic looseness of the TUR, set by the driving $u/w$ and vanishing only near equilibrium. All dependence on the measurement instead lives in the first inequality: the gap between $\langle \dot{S}\rangle_{\rm TUR}$ and this bound is the penalty for the noisy observation, and it closes only in the ideal limit $\epsilon\to 0$, $\mu\to\infty$. Because the driving is held fixed in Fig.~\ref{fig:photon_ep} ($u=2$, $w=1$), the unicyclic TUR bound appears there as a fixed horizontal line just below $\langle \dot{S}\rangle$, and essentially all variation of $\langle \dot{S}\rangle_{\rm TUR}$ across the panels is this measurement penalty.

The separation is conceptual rather than numerical: the measurement penalty is itself larger for more strongly driven motors, which lose more signal to the finite photon rate. In particular $\langle \dot{S}\rangle_{\rm TUR}$ is not monotonic in $\mu$. At any fixed $\epsilon > 0$ it rises to an optimum at an intermediate photon rate and then falls, because the mean photon current saturates at large $\mu$ while its variance keeps growing with the rate of wrong-color emissions. In this respect it behaves much like the Markov estimator; the practical advantages of the TUR estimator are that it is a rigorous lower bound and that it is the more robust of the two at appreciable error probability, whereas the first-order Markov estimator, lacking any analogue of this bound, can approach $\langle \dot{S}\rangle$ only in the near-noiseless regime.

\section{Conclusions}

In conclusion, we have considered two sources of experimental errors that affect thermodynamic inference from single-molecule or single-particle trajectories. One is finite temporal resolution, which results in the system's state being unknown in between infrequent observations. The other is noise, which leads to mislabeling of the observed states of the system. We then explored two classes of approaches. One is based on taking the experimentally measured trajectories $Y(t)$ at face value as if it is the true microscopic observable $X(t)$. An interesting observation is that $Y(t)$ may exhibit non-Markovian dynamics even when the true trajectory $X(t)$ is a Markov process. In this case an improved entropy production estimate can be obtained using non-Markov estimators as well as the thermodynamic-uncertainty-relationship based estimator. Regardless of whether a non-Markov estimator is used, we showed that computing the KL divergence between $Y(t)$ and its time-reversed version provides a lower bound on the entropy production, which generally becomes progressively worse as the noise becomes stronger. The dependence of the estimates on the time resolution is more complicated: In certain scenarios, there is an optimal time resolution where the most accurate estimate is obtained, while in others lower temporal resolution produces better estimates, an effect similar in its origin to the improved entropy estimation using the milestoning method,\cite{Blom2024} where sparser milestones reporting on the system's dynamics results in more accurate thermodynamic inference.

A second approach attempts to recover the statistical properties of the true dynamics $X(t)$ from those of experimental trajectories $Y(t)$. This is only possible when the ``error matrix'' (that is, the matrix of the probabilities $P(Y|X)$ of mislabeling $X$ as $Y$) is invertible. Moreover, given a discrete propagator describing the evolution of $X$ measured at discrete moments of time, the true propagator can be recovered via a matrix logarithm or matrix inversion.\cite{Bauer_2025} Numerically, however, these methods are not necessarily feasible, especially if $P(Y|X)$ is not precisely known, as is often the case in experimental studies.

An alternative to the above methods that was not considered here is statistical inference of the underlying kinetic model.\cite{Gopich2009, Gopich2015, PresseBook, PresseFRET1, PresseFRET2} Indeed, this type of approach has already been used to deduce the nonequilibrium character of the dynamics of molecular machines from single-molecule observations.\cite{Hugel2014,Hugel_HMM} In typical applications, which use hidden Markov models (HMMs), one starts with postulating a kinetic model and then determines the optimal parameters fitting the data. If the assumed model is close enough to the true dynamics, this method may outperform other entropy production estimators even in the absence of experimental noise.\cite{two-track-walk} This strategy, however, is handicapped by the model selection problem, and imposing an inadequate model may lead to physically erroneous inference.\cite{Presse2025HMM} In the context of thermodynamic inference, in particular, this approach is not guaranteed to yield a lower bound and may {\it overestimate} the entropy production.\cite{Song2024FRET} Bayesian non-parametric approaches have been developed to overcome this problem,\cite{PresseBook} albeit often at a high computational cost.

\appendix

\section{Koza method for current cumulants}
\label{app:koza}

\subsection*{Variables used in this and the following appendix}

The expressions in Appendices~\ref{app:koza} and~\ref{app:TUR_proof} are most transparent in the rate-ratio variables $(k, \delta)$ rather than the physical variables $(\mu, \epsilon)$ used in the main text. Here $k$ is the rate at which the motor in state $X$ emits a photon of the \emph{matching} color $Y = X$, and $k\delta$ is the rate at which it emits a photon of each of the two \emph{wrong} colors. So $\delta \geq 0$ measures how often a wrong-color photon is emitted relative to the right-color photon: $\delta = 0$ corresponds to perfect color reporting, and $\delta = 1$ corresponds to all three colors being emitted at equal rates (so the photon color carries no information about the motor state). The total photon rate is $\mu = k(1+2\delta)$ and the per-emission error probability is $\epsilon = \delta/(1+2\delta)$. All results in these appendices transfer to the main-text equations upon the substitution
\begin{equation} \label{eq:variable_map}
k = \mu(1 - 2\epsilon), \qquad \delta = \frac{\epsilon}{1 - 2\epsilon}
\end{equation}
mapping from total photon rate $\mu > 0$ and per-emission error probability $\epsilon \in [0, 1/2)$ to $k > 0$ and $\delta \geq 0$.

\subsection*{The 9-state generator and its tilted form}

The full $(X, Y)$ system is a 9-state continuous-time Markov chain with the master equation given in Eq.~\eqref{eq:nine_state_generator}.
The tilted generator $\mathbb{W}(\lambda)$ is obtained from the continuous-time generator by multiplying each clockwise $Y$-transition rate by $e^{\lambda}$ and each counterclockwise $Y$-transition rate by $e^{-\lambda}$; the $X$-transitions are unaffected.
The generator and its tilted form commute with the cyclic shift $(X, Y) \to (X+1, Y+1) \pmod 3$. The scaled cumulant generating function $\psi(\lambda)$---the largest eigenvalue of $\mathbb{W}(\lambda)$, satisfying $\psi(0) = 0$---therefore lies in the symmetric (cyclic-invariant) sector. In this sector the dynamics reduce to a $3\times 3$ tilted operator
\begin{equation} \label{eq:tilted_3x3}
\mathbb{W}_{\mathrm{red}}(\lambda) = \begin{pmatrix}
\alpha & e^{\lambda}\,k + u & e^{-\lambda}\,k + w \\[2pt]
w + e^{-\lambda}\,k\delta & \beta & u + e^{\lambda}\,k\delta \\[2pt]
u + e^{\lambda}\,k\delta & w + e^{-\lambda}\,k\delta & \beta
\end{pmatrix},
\end{equation}
acting on the basis $\{v_0, v_{+1}, v_{-1}\}$, where $v_d$ is the uniform superposition of the three states with $Y - X \equiv d \pmod 3$ and $\alpha$, $\beta$ are the same escape rates as in Eq.~\eqref{eq:nine_state_generator}. The top eigenvalue of $\mathbb{W}_{\mathrm{red}}(\lambda)$ equals $\psi(\lambda)$.

\subsection*{Cumulants via Koza's method}

Let the characteristic polynomial of $\mathbb{W}_{\mathrm{red}}(\lambda)$ be written as $\det(x\mathbf{I} - \mathbb{W}_{\mathrm{red}}) = x^3 + c_2(\lambda) x^2 + c_1(\lambda) x + c_0(\lambda)$. Since $\psi(0) = 0$, we have $c_0(0) = 0$, and the two non-zero eigenvalues at $\lambda = 0$ have product $c_1(0)$. Expanding $\psi(\lambda)$ to second order via Vieta's relations gives the Koza expressions~\cite{Koza1999,Koza2002}
\begin{align}
\langle j_Y \rangle &= \psi'(0) = -\left.\frac{c_0'(\lambda)}{c_1(\lambda)}\right|_{\lambda=0}, \label{eq:koza_J} \\
\langle \delta j_Y^2 \rangle &= \psi''(0) = -\frac{c_0''(0) + 2\langle j_Y\rangle\bigl[c_1'(0) + \langle j_Y\rangle c_2(0)\bigr]}{c_1(0)}. \label{eq:koza_var}
\end{align}
Evaluating these in the $(k, \delta)$ variables yields the mean current
\begin{equation} \label{eq:mean_current_kdelta}
\langle j_Y \rangle = \frac{k^2(u-w)(1-\delta)^2}{\mathcal{Z}},
\end{equation}
where $\mathcal{Z}$ is the same denominator polynomial as in Eq.~\eqref{eq:mean_current_Y} of the main text, re-expressed in $(k, \delta)$ as
\begin{equation} \label{eq:Z_kdelta}
\mathcal{Z} \equiv 3(u^2 + uw + w^2) + 3k(u+w)(1+2\delta) + k^2(1+2\delta)^2,
\end{equation}
and the variance rate
\begin{equation} \label{eq:variance_kdelta}
\langle \delta j_Y^2 \rangle = \frac{k\,\mathcal{N}}{\mathcal{Z}^3},
\end{equation}
where the polynomial $\mathcal{N}$ is most conveniently written as a polynomial in $k$,
\begin{equation} \label{eq:N_polynomial}
\mathcal{N}(u, w, k, \delta) = \sum_{j=0}^{6} \mathcal{N}_j(u, w, \delta)\,k^j,
\end{equation}
with coefficients
\begin{align}
\mathcal{N}_0 &= 18\,S^3\,(1+2\delta), \notag \\
\mathcal{N}_1 &= 9\,(u+w)\,S^2\,\bigl(5 + 26\delta + 23\delta^2\bigr), \notag \\
\mathcal{N}_2 &= 18\,S\,\Bigl[A_2\,(u^2+w^2) + B_2\,uw\Bigr], \notag \\
\mathcal{N}_3 &= 9\,(u+w)\,\Bigl[A_3\,(u^2+w^2) + 4 B_3\,uw\Bigr], \label{eq:Nj_coeffs} \\
\mathcal{N}_4 &= 6\,(1+2\delta)\,\Bigl[A_4\,(u^2+w^2) + 3 B_4\,uw\Bigr], \notag \\
\mathcal{N}_5 &= (u+w)\,(1+2\delta)^3\,\bigl(1 + 12\delta + 135\delta^2 + 14\delta^3\bigr), \notag \\
\mathcal{N}_6 &= 18\,\delta^2\,(1+2\delta)^4,  \notag
\end{align}
where $S \equiv u^2 + uw + w^2$ and the auxiliary polynomials in $\delta$ alone are
\begin{align}
A_2 &= 3 + 22\delta + 55\delta^2 + 28\delta^3, \nonumber \\
B_2 &= 4 + 40\delta + 97\delta^2 + 48\delta^3, \nonumber \\
A_3 &= 3 + 38\delta + 147\delta^2 + 228\delta^3 + 70\delta^4, \nonumber \\
B_3 &= 1 + 10\delta + 54\delta^2 + 74\delta^3 + 23\delta^4, \nonumber \\
A_4 &= 1 + 20\delta + 99\delta^2 + 170\delta^3 + 34\delta^4, \nonumber \\
B_4 &= 1 + 8\delta + 64\delta^2 + 96\delta^3 + 20\delta^4. \nonumber
\end{align}
Every coefficient of every $\mathcal{N}_j$ is a non-negative monomial in $u, w, \delta$, so $\mathcal{N} > 0$ throughout the parameter domain. This guarantees that the variance rate~\eqref{eq:variance_kdelta} is positive and therefore the TUR ratio is well-defined; it will also do most of the algebraic work in Appendix~\ref{app:TUR_proof}, since the polynomial $P$ used to prove the TUR bound is built directly from $\mathcal{N}$ and $\mathcal{Z}$.

Substituting Eq.~\eqref{eq:variable_map} into Eqs.~\eqref{eq:mean_current_kdelta} and~\eqref{eq:variance_kdelta} reproduces Eqs.~\eqref{eq:mean_current_Y} and~\eqref{eq:variance_rate} of the main text: the $(k, \delta)$ form of $\mathcal{Z}$ in Eq.~\eqref{eq:Z_kdelta} reduces to the noise-independent form in Eq.~\eqref{eq:Z_def}, and the prefactor $(1-\delta)^2$ becomes $(1-3\epsilon)^2$ after collecting the factors of $(1-2\epsilon)$ that the substitution introduces.

\section{Proof of the TUR bound for the coarse-grained photon current}
\label{app:TUR_proof}

We prove Eq.~\eqref{eq:tur_nested}:
\begin{equation}
\frac{2\langle j_Y \rangle^2}{\langle \delta j_Y^2 \rangle} \leq (u-w)\ln\frac{u}{w}
\end{equation}
for all $u, w > 0$ ($u \neq w$), $k > 0$, and $\delta \geq 0$. Equivalently, by the bijection of Eq.~\eqref{eq:variable_map}, this covers all $\mu > 0$ and $\epsilon \in [0, 1/2)$.

The proof proceeds in two steps through the intermediate quantity $2(u-w)^2/(u+w)$.

\subsection*{Step 1: The measurement degrades the signal-to-noise ratio}

We show that
\begin{equation}
\frac{2\langle j_Y \rangle^2}{\langle \delta j_Y^2 \rangle} \leq \frac{2(u-w)^2}{u+w}.
\end{equation}
Using the $(k, \delta)$ form of the cumulants from Eqs.~\eqref{eq:mean_current_kdelta} and~\eqref{eq:variance_kdelta}, the difference between the two sides can be written
\begin{equation} \label{eq:difference_P}
\frac{2(u-w)^2}{u+w} - \frac{2k^3(u-w)^2(1-\delta)^4\,\mathcal{Z}}{\mathcal{N}} = \frac{P(u,w,k,\delta)}{(u+w)\,\mathcal{N}},
\end{equation}
with
\begin{equation} \label{eq:P_def}
P = 2(u-w)^2\bigl[\mathcal{N} - k^3(1-\delta)^4\,\mathcal{Z}\,(u+w)\bigr].
\end{equation}
The denominator $(u+w)\mathcal{N}$ is positive (using $\mathcal{N} > 0$ from Appendix~\ref{app:koza}), so it suffices to show $P \geq 0$.

Expanding in powers of $k$, $P = \sum_{j=0}^{6} P_j(u,w,\delta)\,k^j$. The first piece of Eq.~\eqref{eq:P_def} contributes $2(u-w)^2 \mathcal{N}_j$ to every coefficient, while the second piece, $k^3(1-\delta)^4\mathcal{Z}(u+w)$, contributes only at $j = 3, 4, 5$, since $\mathcal{Z}$ in Eq.~\eqref{eq:Z_kdelta} is a polynomial of degree 2 in $k$ and is multiplied by $k^3$. The four coefficients outside this window are therefore furnished directly by the $\mathcal{N}_j$ of Eq.~\eqref{eq:Nj_coeffs},
\begin{equation} \label{eq:Pj_direct}
P_j = 2(u-w)^2\,\mathcal{N}_j \qquad \text{for } j \in \{0, 1, 2, 6\},
\end{equation}
each non-negative because the $\mathcal{N}_j$ have only non-negative monomials in $u, w, \delta$. For $j \in \{3, 4, 5\}$, the correction from the second piece gives
\begin{equation} \label{eq:Pj_corrected}
P_j = 2(u-w)^2\bigl[\mathcal{N}_j - \mathcal{C}_j\bigr] \qquad \text{for } j \in \{3, 4, 5\},
\end{equation}
where the correction polynomials
\begin{equation} \label{eq:Cj_def}
\begin{aligned}
\mathcal{C}_3 &= 3 S\,(u+w)\,(1-\delta)^4, \\
\mathcal{C}_4 &= 3\,(u+w)^2\,(1+2\delta)\,(1-\delta)^4, \\
\mathcal{C}_5 &= (u+w)\,(1+2\delta)^2\,(1-\delta)^4
\end{aligned}
\end{equation}
come from the $k^0$, $k^1$, and $k^2$ pieces of $\mathcal{Z}$ in Eq.~\eqref{eq:Z_kdelta}, respectively. Each $\mathcal{N}_j - \mathcal{C}_j$ written in this form involves a difference, so its non-negativity is not obvious. Expanding and collecting by monomials in $u, w$, however, exposes the structure:
\begin{equation} \label{eq:Q3_explicit}
\begin{aligned}
\mathcal{N}_3 - \mathcal{C}_3 &= (u^3 + w^3)\,Q_3 + (u^2 w + u w^2)\,R_3, \\
\mathcal{N}_4 - \mathcal{C}_4 &= (u^2 + w^2)\,Q_4 + uw\,R_4, \\
\mathcal{N}_5 - \mathcal{C}_5 &= (u + w)\,Q_5,
\end{aligned}
\end{equation}
where the Q and R terms are polynomials in $\delta$
\begin{align}
Q_3 &= 24 + 354\delta + 1305\delta^2 + 2064\delta^3 + 627\delta^4, \notag \\
R_3 &= 57 + 726\delta + 3231\delta^2 + 4740\delta^3 + 1452\delta^4, \notag \\
Q_4 &= 3 + 138\delta + 840\delta^2 + 2184\delta^3 + 2265\delta^4 + 402\delta^5, \notag \\
R_4 &= 12 + 192\delta + 1452\delta^2 + 3984\delta^3 + 3858\delta^4 + 708\delta^5, \notag \\
Q_5 &= 18\delta + 225\delta^2 + 972\delta^3 + 1791\delta^4 + 1260\delta^5 + 108\delta^6 \notag
\end{align}
with only positive integer coefficient.
Therefore every $\mathcal{N}_j - \mathcal{C}_j \geq 0$, every $P_j \geq 0$, every $P_j k^j \geq 0$, and $P \geq 0$.

\subsection*{Step 2: The unicyclic TUR}

The inequality $2(u-w)^2/(u+w) \leq (u-w)\ln(u/w)$ is the standard TUR for the 3-state symmetric ring. Assuming $u > w > 0$ and dividing by $u - w > 0$, this reduces to
\begin{equation}
\frac{2(r-1)}{r+1} \leq \ln r, \qquad r \equiv u/w > 1.
\end{equation}
Defining $f(r) = \ln r - 2(r-1)/(r+1)$, we have $f(1) = 0$ and $f'(r) = (r-1)^2/[r(r+1)^2] \geq 0$, so $f(r) \geq 0$ for all $r \geq 1$. The case $r < 1$ follows by symmetry.

Combining Steps 1 and 2 completes the proof. Physically, Step~1 is an information-processing inequality---the stochastic photon emission can only add noise---and Step~2 is the thermodynamic bound on the motor's intrinsic signal-to-noise ratio.

\begin{acknowledgements}
 This work was initiated and performed partly at the Aspen Center for Physics, supported by National Science Foundation grant PHY-2210452. The research is also supported by the Welch Foundation award C-1559 (to ABK), as well as by National Science Foundation awards (CHE-2400424 to DEM and MCB-2204402 to OAI) and by the Camille Dreyfus Teacher-Scholar Award to TRG. OAI and ABK are also partially supported by the Center for Theoretical Biological Physics sponsored by the National Science Foundation (PHY-2019745). Discussions with Kristian Blom, J\"org Enderlein, and Matthias Kr\"uger are gratefully acknowledged. 
\end{acknowledgements}

\bibliography{bibl_currated.bib}

\end{document}